# KHAIT: K-9 Handler Artificial Intelligence Teaming for Collaborative Sensemaking

Matthew Wilchek
Department of Computer Science
Virginia Polytechnic Institute and
State University (Virginia Tech)
Arlington, Virginia, USA
mwilchek@vt.edu

Linhan Wang
Department of Computer Science
Virginia Polytechnic Institute and
State University (Virginia Tech)
Arlington, Virginia, USA
linhan@vt.edu

Sally Dickinson
School of Animal Sciences
Virginia Polytechnic Institute and
State University (Virginia Tech)
Blacksburg, Virginia, USA
sallyd@vt.edu

Erica Feuerbacher
School of Animal Sciences
Virginia Polytechnic Institute and
State University (Virginia Tech)
Blacksburg, Virginia, USA
enf007@vt.edu

Kurt Luther
Department of Computer Science
Virginia Polytechnic Institute and
State University (Virginia Tech)
Arlington, Virginia, USA
kluther@vt.edu

Feras A. Batarseh
Department of Biological Systems
Virginia Polytechnic Institute and
State University (Virginia Tech)
Arlington, Virginia, USA
batarseh@vt.edu

## Abstract

In urban search and rescue (USAR) operations, communication between handlers and specially trained canines is crucial but often complicated by challenging environments and the specific behaviors canines are trained to exhibit when detecting a person. Since a USAR canine often works out of sight of the handler, the handler lacks awareness of the canine's location and situation, known as the "sensemaking gap." In this paper, we propose KHAIT, a novel approach to close the sensemaking gap and enhance USAR effectiveness by integrating object detection-based Artificial Intelligence (AI) and Augmented Reality (AR). Equipped with AI-powered cameras, edge computing, and AR headsets, KHAIT enables precise and rapid object detection from a canine's perspective, improving survivor localization. We evaluate this approach in a real-world USAR environment, demonstrating an average survival allocation time decrease of 22%, enhancing the speed and accuracy of operations.

## 1 Introduction

In the wake of natural and manmade disasters (e.g., earthquakes, wars), *sensemaking*, a decision-making process involving plausible explanations for human actions, becomes vital to rescue missions [18, 61]. The complex effort to search and locate trapped survivors from collapsed structures or debris is a race against time. Rescuers who conduct this life-saving activity specialize in urban search and rescue (USAR). USAR teams are composed of highly trained professionals from various fields, including: technical rescue specialist, search specialists, medical, engineers, planning/technical info staff, communications, hazmat, and canine. Canine search teams are trained to operate independently in austere and challenging environments, where they use their sense of smell to isolate the highest concentration of human odor against a backdrop of non-target odors. These specialized canines are trained either to detect the live human odor of survivors (live-find) or the odor of human remains of the deceased (HRD). Live-find and HRD canines are deployed at different mission stages, with live-find being an immediate, high-risk, and time-sensitive operation.

When a canine detects the highest concentration of odor from a survivor buried in the rubble, the dog emits a trained final response (TFR) [11]. This specific behavior is under stimulus control of the target odor the dog has detected. For instance, a live-find USAR dog is trained to bark continuously in the location of the highest concentration of live human odor. Certification standards require this signaling to be temporal to the canine determining the highest odor concentration location [17]. Delays in rescue operations arise from the rescuers' need to determine access routes to and the precise location of the survivor, relative to where the canine indicates the strongest scent, emanating from the rubble. While the term "handler" refers to a specific type of rescuer who works directly with the canine, other rescuers may assist a handler in navigating and accessing the indicated location.

Handlers must rapidly make sense of the complex environment, often contending with navigational challenges, obstructed lines of sight, and restricted access due to the instability of collapsed structures. In these situations, understanding where the dog has detected the strongest scent and how they arrived there can be critical, as it guides rescuers to the most likely location of the survivor, increasing the handler's situational awareness and allowing for greater risk reduction. Global Positioning Satellite (GPS) navigation can be ineffective due to lack of signal or failure to perceive the depth of the survivor's location [58].

Augmented Reality (AR) has increasingly been recognized for its potential to enhance sensemaking among emergency response professionals by providing real-time, contextual information that supports decision-making in complex environments [5, 33, 67]. Traditionally, AR applications have focused on augmenting human sensory perception to navigate and interpret emergency scenes more effectively [32, 36, 53]. However, the potential of integrating sensor-based positioning, and simultaneous localization and mapping (SLAM) based AR with data from canines, which play critical roles in USAR operations, represents a novel and largely untapped opportunity.

Few existing canine harnesses can capture invaluable data such as scent detection, temperature, and the physical movements of canines [29]. While Artificial Intelligence (AI) has primarily been incorporated into smart canine collars for health monitoring purposes [9, 22, 26], its application in enhancing operational capabilities has been less explored. Various smart harnesses have been developed to monitor physiological and environmental data [6, 42, 57], yet the integration of AI to support real-time decision-making by handlers remains relatively unexamined [7, 39, 50]. This gap highlights an opportunity for more advanced AI-driven systems that not only track health metrics but also augment AI-teaming between a canine and their handler in critical situations such as search and rescue operations.

K-9 Handler Artificial Intelligence Teaming (KHAIT) is our approach that extends the capabilities of AR and AI to enhance USAR operations. KHAIT integrates a smart AI-enabled canine harness with a multi-user AR application, enabling the translation of a canine's detected data into visual cues on the handler's AR display. Specifically, handlers see a real-time video feed from the canine's perspective, enhanced by AR overlays. This video is stabilized and positioned within the user's physical space using computer vision techniques for markerless tracking and SLAM [10]. Critical elements within the video, such as potential survivors or hazards, are highlighted by bounding boxes generated by the YOLOv8l AI model, which has demonstrated a mean Average Precision (mAP) of 52.9 on the COCO dataset [27]. This integration facilitates real-time visualization and AI-enabled sensemaking of camera data, effectively incorporating the human-in-the-loop (HITL) concept by enabling handlers to interact with and respond to the information dynamically.

Capitalizing on the canine's acute sensory perceptions, KHAIT supports the handler's situational awareness and enables scalable object detection sharing across various distances and obstacles. Additionally, this modality allows for the continuous collection of diverse data from a canine harness, crucial for refining AI algorithms and augmenting AI-teaming between detection canines and their handlers in disaster response scenarios. We define two research questions (RQ) that shape the focus of this study:

- **RQ1:** Can AI-powered cameras and AR headsets enhance the accuracy and efficiency of canine handlers in localizing survivors during USAR operations, and what measurable benefits do these technologies provide in real-world search scenarios?
- **RQ2:** From a usability standpoint, how does integrating AR and AI technologies into canine harnesses and human-operated systems affect the ease of interpretation and interaction for both SAR canines and their handlers during searches?

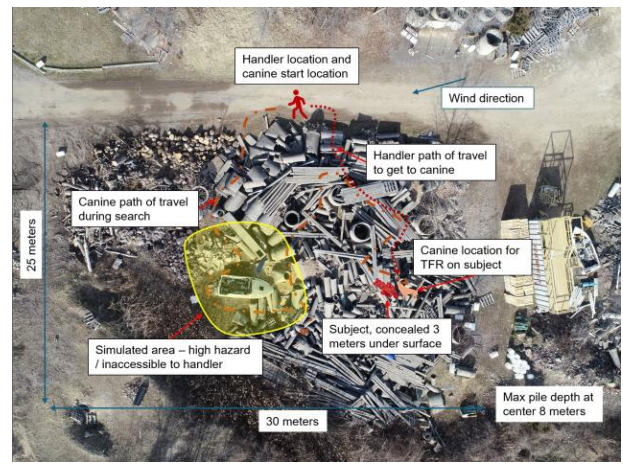

**Figure 1: Search scenario at a USAR training facility illustrating the difference between using the KHAIT system and traditional search methods. After releasing the canine, the handler remains at the start location, simulating a restricted access site. Without KHAIT, the handler attempted to follow the canine's initial path through hazardous rubble to the TFR location, encountering impassable obstacles. With KHAIT, the handler had spatial awareness of the rubble pile and could quickly formulate alternate routes to access the canine location.**

Our evaluation of KHAIT included two studies with human participants and canines to understand its sensemaking affordances in the context of a simulated rescue operation. The first was a pilot study that collected preliminary results from fellow graduate students and one of their pet dogs. The second study consisted of trained SAR rescuers and USAR canines at an official USAR training facility as depicted in Figure 1. The preliminary results from the pilot study showed promising technical reliability of the system and enhanced sensemaking for tasks involving the identification and localization of simulated survivors needing help. The real-world study with trained USAR rescuers and canines demonstrated a 22% reduction in time to locate simulated survivors. Participants also rated KHAIT's usability as acceptable using two validated questionnaires post-trial. Our main contributions can be summarized as follows:

- We introduce KHAIT, a novel shared perception system that integrates a smart AI-powered canine harness with an AR device, enhancing the sensemaking process between canines and their handlers.
- We are the first to explore data visualization using SLAM-based AR from a canine's perspective, examining how this capability can assist handlers in real-time decision-making.
- We evaluate the effectiveness of KHAIT within a real-world urban search and rescue environment involving four trained rescue canines and five first responders, demonstrating a significant reduction in task completion times.

## 2 Related Work

HCI and ACI are two fields that increasingly intersect in the design of interactive systems. HCI focuses on developing systems that enhance human capabilities, emphasizing usability, accessibility, and user experience to ensure that technology amplifies human control and meets user needs [55]. ACI extends these principles to non-human animals, advocating for technologies that respect animal well-being, autonomy, and engagement [14]. The integration of these two fields is particularly pertinent in environments where humans and animals collaborate closely, such as in USAR.

In the increasingly complex interactions between humans, animals, and technology, AI can act as an intermediary that enriches the interaction between HCI and ACI. AI's capabilities to process and analyze big data in real-time can significantly enhance HCI and ACI; making systems more intuitive to humans' and animals' needs and behaviors. In the context of USAR, AI-driven technologies can interpret complex animal signals [16, 23], such as those from search dogs, and translate them into actionable data for humans. This integration not only amplifies the natural capabilities of search dogs by providing handlers with interpretive tools, but also ensures that the technology adapts to the animal's behavioral cues to respect their autonomy and well-being.

### 2.1 AR for Search and Rescue

In USAR operations, the integration of AR has shown potential in enhancing human sensemaking capabilities [12, 25]. Innovations such as ruggedized HoloLens adaptations like the C-THRU helmet, IVAS, and MARS underscore AR's role in improving operational efficiency and situational awareness [31, 46, 54]. From a research perspective, Smith et al.'s 2022 software for HoloLens aids in bridge inspections through annotations and spatial markers [56], while Guan et al. introduced an object detection app that overlays 3D bounding boxes on detections [21]. These advancements, however, primarily focus on enhancing individual user experiences and do not fully address collaborative scenarios involving multiple users or integrate non-human data sources effectively [1, 34, 65].

In recent years, wearable mobile interfaces have been developed to enhance sensemaking for SAR dog handlers [66]. Most similar to KHAIT, RescueGlass uses Google Glass and a mobile phone to display the location information of a canine and other rescuers [49]. However, RescueGlass relies on wireless connectivity, is not completely hands-free (given a mobile phone is involved), is not AI-driven, and the AR display isn't SLAM enabled, which limits interactive UI functionalities.

Current research largely explores AI-enhanced AR systems in single-user contexts, with less attention given to multi-user environments that are typical in USAR operations [60]. However, recent research has witnessed a surge in investigating the application of human-in-the-loop (HITL) learning for object detection on AR devices, aiming to enhance the process of sensemaking. The involvement of human users through HITL learning has proven to enhance object detection performance [62].

However, these initiatives often do not extend to collaborative, distributed multi-user scenarios, nor do they adequately address the integration of data from canine partners, which is crucial for comprehensive sensemaking in SAR [24]. Furthermore, the computation for object detection often occurs remotely on a web server rather than on the physical HoloLens device or an alternative edge device. This raises concerns regarding operating in real-time environments with poor or insecure signals. In Section 3, we discuss how the system design of KHAIT addresses these gaps by integrating multi-user AR functionality and canine-derived data into an integrated system of systems with reliable supporting infrastructure.

### 2.2 Smart Canine Harnesses for Search and Rescue

While HCI has seen extensive application in SAR, the specific integration of ACI, particularly using canine smart harnesses for real-time sensemaking, remains as an active research topic. Tran et al. [59] proposed an early example of a canine harness to enhance sensemaking in USAR applications that leveraged the Microsoft Kinect. Similarly, Alcaidinho et al. [2] showcase wearable technology for canine units, featuring sensors and GPS on a harness to relay critical events, such as explosive detection, to officers through a ruggedized cellphone.

Pai et al. [40] introduce another smart guide canine harness utilizing AI edge computing for assisting visually impaired individuals, integrating image recognition and navigation technologies. Conversely, Kasnesis et al. [29] leverage a different wearable-based approach for SAR canines, utilizing deep Convolutional Neural Networks (CNN) to detect activities and barking in real-time by analyzing data from inertial sensors and microphones. The system shows promising results in improving communication between SAR canines and their handlers by providing immediate feedback on their status and findings. While these systems are notable for their contributions to wearable tech and activity detection, their designs may not be practical in real-world USAR environments. These environments require systems that do not depend on consistent internet connectivity for communication and should ideally support more hands-free interaction, rather than requiring a mobile device to manage the user interface of a heads-up display.

An aspect that has received less attention is the canine's behavior and well-being when using a smart harness. Farrell et al. [15] propose several values and principles researchers should consider when integrating ACI technology within the assistance canine training (ACT) industry. Understanding the canine's physiological response to the environment can guide handlers in making difficult decisions, such as when to give the canine a break when they are unduly stressed or suffering from heat exhaustion. Zhao et al. and Pelgrim et al., discuss approaches to monitor such activity using smart eyewear that a canine can wear, but the data is uploaded to a cloud environment to analyze later with AI [41, 68]. Physiological markers such as heart rate, heart rate variability, and temperature could provide insight into how the environment, in conjunction with work-rest cycles, affects the canine and subsequently affects their detection capabilities. There are a variety of smart harnesses and collars available on the market [19, 37, 47], and the integration of this health data from the canine is a long-term goal of our work. These research gaps highlight a critical need for a system that collects data from canines not only safely but also seamlessly integrates it into AR interfaces used by human rescuers. Such integration would represent a further step forward in operational collaboration between humans and canines [20, 51], leveraging the strengths of both to improve team sensemaking and decision-making in high-stakes environments.

## 3 Methods

### 3.1 User Scenario

As mentioned above, traditional search methods in partially collapsed structures involve canines locating potential survivors and signaling their handlers. The handler must navigate the debris to reach the canine's location. When a canine emits its TFR with KHAIT, its location is transmitted to the handler's wearable AR device. This information enables the handler to navigate to the canine and survivor's location swiftly and efficiently, guided by a hologram video feed visible through the AR headset. The handler can then project a marker anchored to the precise location of the canine or survivor via SLAM, which remains visible even through physical obstructions, and communicate the hologram marker to the rescue team. Finally, all first responders reach the marker, triage the survivor, and/or request specific types of support as needed.

### 3.2 Hardware Details

The integration of HCI and ACI was crucial for the hardware design. Our approach focused on creating a usable AR interface for a canine handler while ensuring the comfort and safety of the canine wearing a new smart AI harness. Prior research indicated that carrying more than 10% of a canine's body weight can be excessively burdensome [28, 52], so weight considerations played an important role in the harness design.

The first canine to test our harness was a companion dog, a female husky named Luna, age 3, weighing 20.5 kg and owned by one of the authors of this paper. We selected lightweight, efficient components to balance the need for powerful computing and long testing durations against the risk of overburdening the canine. The total weight of the wearable harness is 2 kg. As illustrated in Figure 2, it includes an NVIDIA Jetson Orin Nano (8GB), which offers substantial computational power for its size, supporting advanced AI tasks without the heft of larger units. The camera is a CMOS 4K Autofocus model, which provides high-resolution imaging critical for effective AR visualization yet is small and light enough to fit comfortably on the harness. The 27000mAh portable power

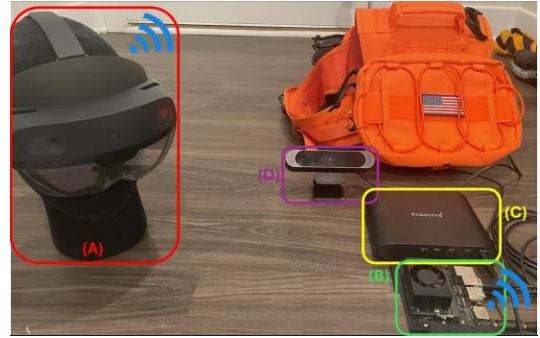

**Figure 2: KHAIT Hardware Components. (A) HoloLens-2 for canine handler. (B) AI edge computing module. (C) Power bank. (D) CMOS 4k autofocus camera.**

bank, while providing extended testing duration, was the heaviest component.

The Jetson Orin Nano [38] processes an advanced embedded object detection algorithm, YOLOv8l. The object detections are visualized with bounding boxes and relayed directly to an AR headset worn by the canine handler through socket programming, over a secure local network. To relay data communication between the Orin Nano and the AR headset, we created a local mesh network using two nodes of the ASUS ROG Rapture GT6 Tri-band WiFi 6 that covered about 5,800 ft$^2$.

### 3.3 AI Object Detection

The NVIDIA Orin Nano is used to transmit a real-time video feed from the canine's perspective, with object detection bounding boxes overlaid to highlight detected objects. The Orin Nano acts as a server that hosts the video feed so that the canine handler wearing the HoloLens-2 can connect to it via the private mesh network. The AI model used for object detections is YOLOv8l from Ultralytics [48]. Since the HoloLens-2 is SLAM-based, the video feed can be spatially anchored in a user-defined position, as shown in Figure 3. The user can choose to have the video feed move with them based on head-tracking or remain static in a stationary position. The user can also scale the video feed window to whatever size needed. We found these features especially helpful for rescuers as they moved around or investigated an area, as will be discussed in Section 7.

### 3.4 Software Architecture

Our system incorporates various hardware components as depicted in Figure 4. The software architecture was designed to maximize connectivity and data integration across these platforms through a local private mesh network. We utilize Unity, a software development tool for AR, and the Mixed Reality Toolkit 3 (MRTK) for developing the 3D User Interface on AR devices. The backend server on the edge device employs the Flask Python network framework, complemented by the Redis package for its high usability and performance.

Besides the real-time video feed with AI detections, we integrated our software with another AR module called Ajna that incorporates advanced localization capabilities to enhance target localization

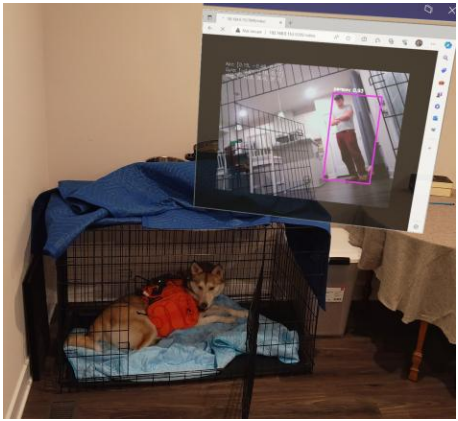

**Figure 3: Screenshot of the KHAIT prototype from a handler's perspective. The live video and AI detections captured by the camera equipped on Luna (canine) are projected onto the handler's AR device, allowing the handler to become familiar with the environment.**

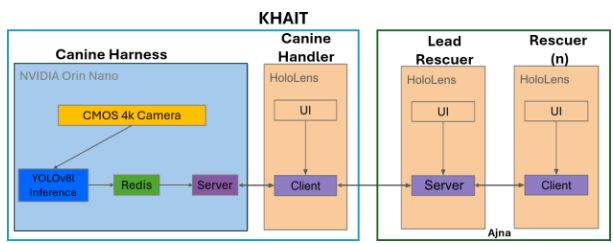

**Figure 4: Software Architecture of KHAIT. The canine harness integrates various components linked to a server via Redis for real-time data handling from a canine. YOLOv8l is used for object detection, communicating results to the Canine Handler's HoloLens through the Client-Server model. Due to Ajna's design, KHAIT can be extended to multiple rescuers to interact with the detected data simultaneously, though this capability depends on the network infrastructure's robustness.**

[63]. Once the canine handler reaches the location the canine led them to; the handler can then project an AR marker anchored to the physical location of the survivor as demonstrated in Figure 5. The marker can be sent to other rescuers nearby, who would also be wearing a HoloLens-2 and can see the visualization through walls, ceilings, or other objects with minimal spatial drift. In a USAR context, this design choice intends to improve the user's ability to perceive the level of depth someone may be in a rubble pile or building. This approach ensures that our system is robust, offering reliable backup solutions for precise localization and object detection in complex environments.

## 4 Pilot Study

To test the performance of KHAIT initially, we designed a short pilot study to provide preliminary insights. It involved four graduate students acting as rescuers, each tasked with locating a "survivor"

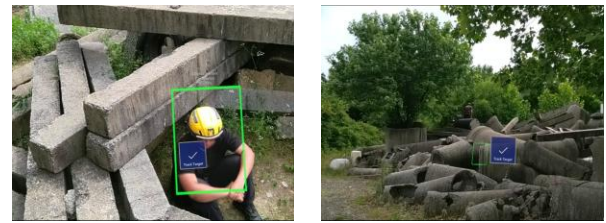

(a) Canine Handler Sends AR Marker to Other Rescuer  (b) Other Rescuer Sees Survivor Location through Rubble

**Figure 5: AR Object Detection Results. (a) AR visualization showing a survivor's location is marked by the canine handler and sent to other rescuers. (b) A second rescuer sees the survivor's location through the rubble pile, demonstrating the system's precision in maintaining the spatial accuracy of virtual objects through physical barriers.**

actor within a four-level, 2,300 ft² residential building to simulate a search area with multiple compartments and the canine (Luna) operating in a remote location to a rescuer. The building included multiple rooms, staircases, and obstructions to mimic surprise hindrances in navigating to the survivor's location.

Students who engaged in the search tasks without the AR headsets faced several challenges that impeded their efficiency. Without real-time visual guidance, each student inadvertently entered one or two rooms where Luna, the search dog, had not been, thus wasting critical time. Additionally, obstacles such as chairs and bins were placed randomly to simulate barriers encountered in real USAR scenarios, which slowed participants down and posed safety risks as they attempted to navigate the multi-level building and access the canine quickly.

In contrast, using the KHAIT system with AR headsets changed how the students approached the search task. With AR, participants received live visual feeds that tracked Luna's movements through the building. This real-time data stream allowed the rescuers to see exactly which paths Luna had taken, avoiding rooms she had not entered. It also allowed participants to foresee obstacles through the AR feed, enabling them to prepare and navigate more effectively compared to trials without AR support.

The pilot study revealed how several factors directly influenced the performance of the KHAIT system. Participants found the user interface intuitive and easy to use, with clear visual markers and a logical layout, which allowed them to focus on the task rather than the technology. This positive reception highlights the importance of having an interface that provides necessary information without distracting rescuers, especially in high-stress scenarios where time is critical. From a system integration perspective, all the technical components worked with no issues and communicated effectively between each device over the local mesh network.

## 5 Evaluation

Based on our preliminary results from the pilot study, we designed a more comprehensive experimental evaluation of the KHAIT system, using a design within subjects to assess its effectiveness in real-world USAR scenarios. We conducted search exercises at an official

**Table 1: Canines Participating in SAR Studies with KHAIT**

| Canine | Breed (Name) | Gender | Age | Weight (KG) | Study Used | Harness Compatibility | Accepted in User Study |
|---|---|---|---|---|---|---|---|
| 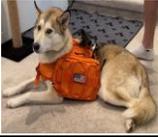 | Siberian Husky (Luna) | Female | 3 | 20.4 | Pilot Study | ✓ | ✓ |
| 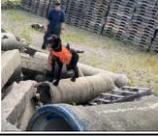 | German Wirehaired Pointer (Peat) | Male | 3 | 30.3 | User Study | ✓ | ✓ |
| 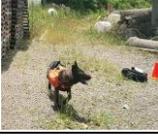 | German Shepherd (Memphis) | Female | 2 | 33.5 | User Study | ✓ | ✓ |
| 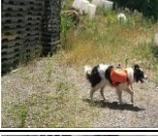 | Border Collie (Loki) | Male | 9 | 15.8 | User Study | ✗ | ✗ |
| 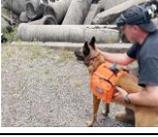 | Belgian Malinois (Val) | Female | 6 | 31.3 | User Study | ✓ | ✗ |

USAR training facility featuring partially collapsed structures and concrete debris. The evaluation involved five professional USAR rescuers and four trained USAR rescue canines. The design mirrored the pilot study, with each rescuer and canine team conducting similar trials to assess the system under more challenging and realistic conditions. The evaluation was approved by our university's Internal Review Board (IRB) (#24-389, #24-569) and Institutional Animal Care and Use Committee (IACUC) (#24-102).

Each rescuer, designated as R1 through R$n$, participated in two trials, one utilizing the KHAIT system (experimental condition) and another without it (control/baseline condition). The order of the trials was counterbalanced to mitigate potential learning effects and stress, ensuring that each trial presented similar technical difficulty and distance of access from the start point to the survivor. Additionally, each scenario included a simulated hazard or no-go zone, unknown to the handler at the start, to further test the efficacy of the KHAIT system under comparable and controlled conditions. This method ensured that all rescuers experienced both conditions under consistent circumstances, allowing for a valid comparison of the results.

The evaluation focused on measuring KHAIT's efficiency and usability. We operationalized efficiency as the time it took for rescuers to locate the survivor following detection by the canine and handler. We hypothesized that rescuers equipped with the KHAIT system would achieve faster times due to improved situational awareness provided by the AR interface, which displayed real-time video and data overlays from the canine's perspective, aiding in navigating complex layouts and identifying direct routes to the survivor. We recorded the time to locate the survivor in each trial, comparing the durations between KHAIT-assisted and traditional search methods reliant solely on canine auditory cues. This data helped us quantify the time efficiency gained by integrating AR technology in SAR operations.

To evaluate usability, we collected subjective experience data from participants regarding the usability and practicality of KHAIT using two validated questionnaires, the System Usability Scale (SUS) [4] and Mixed Reality Concerns (MRC) Questionnaire [30]. This included their perceptions of how effectively the AR interface and the data integration from the canine harness aided their search efforts. Questionnaires were administered post-trial to capture detailed reactions and suggestions from the participants and handlers, focusing on the system's interface, the comfort of the harness on the canines, and the overall impact on their task performance. Each rescuer received a 10-minute introduction by the principal investigator in using the HoloLens-2 and the AR interface before conducting the trials. This approach provided an initial understanding of the potential benefits and areas for improvement in the KHAIT system, ensuring it could be refined to serve better human rescuers and their canine partners in life-saving SAR operations.

## 5.1 Experimental Constraints

The nature of our evaluation within USAR operations introduces inherent variability that makes each simulated trial challenging to replicate precisely. Due to the dynamic and unpredictable real-world conditions simulated in the trials, creating a consistent baseline for comparison across different runs is not feasible. Factors such as the initial position of the canine relative to the debris pile, the survivor's location within that pile, the presence of physical obstacles, and the specific rescue canine involved in each trial introduce many variations. These variations mean that the quantitative measures, particularly the time to locate survivors, could require thousands of replicated trials to produce statistically reliable data. However, the primary value of the user study lies in the qualitative feedback regarding the enhancement of situational awareness provided by the integration of our system. This feedback from subject matter experts and trained rescue canines offers valuable insights into how sensemaking and decision-making processes can be improved in actual SAR scenarios. Such insights provide an understanding of the practical applications and benefits of our technology in enhancing the effectiveness of rescue operations beyond the quantification of time savings.

## 5.2 Animal Welfare

Ensuring the welfare of canines participating in SAR was a primary concern, especially when integrating new technologies. The canines used in our studies were selected based on their training and natural aptitude for working in high-stress environments typical of disaster scenes. Before the trials, all canines underwent a period of habituation with the KHAIT harness to ensure safe fitting and comfort and to minimize the distraction caused by the novel equipment.

Four trained rescue canines were used to evaluate KHAIT. Table 1 shows each dog's traits and, ultimately, which ones were used for the user study. The canine Loki could not habituate appropriately with the harness prior to the first trial as the harness straps did not tighten sufficiently for the harness to secure on his body. Additionally, the harness-to-canine weight ratio was maximized. Canine Val appeared to be habituated to the KHAIT harness; however, her behavior became less independent during searches, suggesting she would need more time to become comfortable working independently with KHAIT. Therefore, canines Loki and Val were omitted from further operational trials.

The handlers were skilled at recognizing signs of stress and discomfort in their canines, using a monitoring protocol that included behavioral observations and potential opt-in/opt-out signals. This protocol allowed canines the flexibility to continue or pause their participation in the search tasks, ensuring that their participation was as stress-free as possible. This approach ensured that the training and deployment processes were optimized for the canines' comfort, thereby minimizing the likelihood of them opting out.

## 6 Results

This section presents the results from a real-world user and animal study that evaluates the KHAIT system. The user study offers a detailed analysis of the data gathered during the SAR trials, providing insights into the effectiveness of the KHAIT system in enhancing search operations through AR and AI integration. Additionally, we examine observations from an animal study, which assesses the comfort and behavioral responses of our canine equipped with the KHAIT harness, aiming to evaluate the overall integration of the system from the perspective of future trained SAR canines.

### 6.1 Efficiency

The deployment of the KHAIT system in the USAR scenarios varied regarding operational efficiency. Despite the varied performance across individual trials, there was an overall average decrease of 22% in time when rescuers utilized KHAIT compared to operations without it as shown in Table 2. Although individual outcomes varied, this average improvement points to the system's potential benefits. We also recorded the times it took a rescuer to find a survivor from the TFR of a canine. In trials where KHAIT improved response times, such as those involving Rescuer 3 and Rescuer 5, the system greatly enhanced the rescuers' situational awareness. For these rescuers, providing real-time visual data from the canine's perspective allowed for more efficient navigation of the rubble, resulting in time reductions of 57.35% and 43.94%, respectively.

However, there were also trials where KHAIT did not improve efficiency. For example, Rescuer 2 experienced a slower performance with KHAIT, which can be attributed to working with Memphis, a relatively new trained rescue canine that required more time to search and track a scent through the rubble pile. The inconsistency across trials illustrates that while KHAIT provides technological advantages, its effectiveness can vary depending on specific rescue scenarios, canines used, and individual user interaction with the system. This highlights the need for further system refinement to enhance usability under diverse conditions and to minimize cognitive demands on users. Despite these variances, the overall average improvement of 22% in time efficiency demonstrated KHAIT's potential to enhance USAR operations. Further research with a larger sample size or additional metrics might provide a more comprehensive understanding of KHAIT's effectiveness.

**Table 2: Comparative Analysis of SAR Times with and without KHAIT**

| Rescuer (R) | Canine | Time w/o K. (s) | Time w/ K. (s) | Post-TFR w/o K. (s) | Post-TFR w/ K. (s) | Abs. Reduced (s) | % Reduced |
|---|---|---|---|---|---|---|---|
| R1 | Peat | 218 | 277 | 55 | 41 | 14 | **25.45%** |
| R2 | Memphis | 167 | 115 | 34 | 43 | -9 | **-26.47%** |
| R3 | Peat | 216 | 67 | 68 | 29 | 39 | **57.35%** |
| R4 | Peat | 62 | 63 | 33 | 30 | 3 | **9.09%** |
| R5 | Peat | 452 | 217 | 66 | 37 | 29 | **43.94%** |

## 6.2 Usability

After each rescuer's trial of the KHAIT system was completed, the participants provided open-ended feedback about what they thought of the trials. Then, they answered two questionnaires: the SUS and the MRC. The data collected from these questionnaires were systematically analyzed to identify common themes and participant perceptions.

All participants expressed positive sentiments toward using KHAIT and hoped the technology could be refined for operational use. For example, Rescuer 1 stated, *"The ability to observe AI detections from a search can give us clues of a survivor's surroundings. I think this can not only be helpful for urban SAR, but wilderness as well."* Rescuer 2, the lower-performing team in efficiency noted that KHAIT substantially assisted in navigating and determining where to go once Memphis was in TFR, as shown by their high SUS questionnaire responses in Figure 6. Rescuer 2 remarked, *"This is sorcery and science!"* They added, *"I had a pretty good idea where the victim was."* They explained further, *"From an operational perspective, it can take us a while to navigate where our canine went."* Finally, they noted, *"KHAIT gives us a better idea of what to expect."*

Other participants highlighted how KHAIT improved their situational awareness. Rescuer 3 stated, *"When Peat was working and not looking down, I knew where he was working. I wouldn't have known from that perspective the spots he was focusing on without the system."* Rescuer 4 noted, *"I was able to see where the canine was investigating and landmarks that directed me where to go after the TFR."* Rescuer 5, who had the most experience in SAR, provided operational insights into how KHAIT could be used for future SAR and said, *"I truly think the niche with KHAIT is the extreme vigilance towards an area without the TFR, especially in more difficult or inaccessible areas we cannot easily get searchers off the pile, then we can send the engineers and rescuers up there."*

Based on the participant's feedback, we can note their appreciation for the AI-enhanced video feed delivered in AR from the canine's perspective. This feature allowed handlers to preemptively see and circumvent potential obstacles or no-go zones set up by evaluators, thus facilitating more efficient navigation routes and enhanced situational awareness.

### 6.2.1 SUS Results.
The SUS provides a score ranging from 0 to 100, where higher scores indicate better usability. Generally, a score above 68 is considered above average and suggests good usability, scores around 50 are considered average, indicating mediocre usability, and scores below 50 are considered below average, reflecting poor usability. Below is a summary of the responses and overall score by each participant in Figure 6:

All SUS questions used can be found in Appendix A. By following the SUS calculation methods of Bangor et al, the average score for KHAIT was rated as **76.5**; which maps to a grade of B or good [3]. The following is an overview of the key insights drawn from the rescuers' feedback on KHAIT's usability and their respective overall scores.

- **Rescuer 1:** Showed ambivalence towards frequent usage and felt the system was complex and cumbersome, leading to a moderate overall impression. **Overall score: 62.5.**

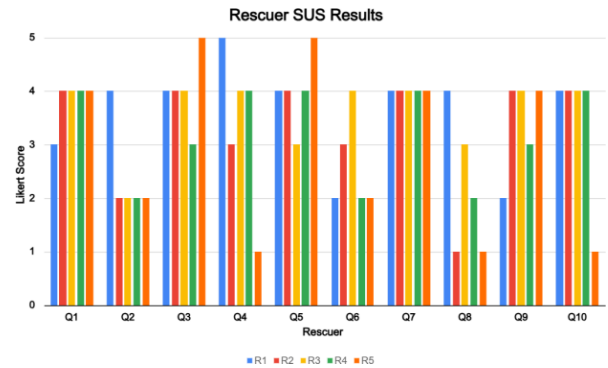

**Figure 6: Rescuer SUS Results**

- **Rescuer 2:** Responded positively, finding the system straightforward and user-friendly, which reflected a high level of comfort with the technology. **Overall score: 80.**
- **Rescuer 3:** Comfortable with regular use and appreciated the system's ease of use, though noted some aspects of inconsistency that impacted their experience. **Overall score: 67.5.**
- **Rescuer 4:** Generally positive, seeing the system as easy to use but indicating a need for technical support, suggesting a balance between usability and dependency. **Overall score: 72.5.**
- **Rescuer 5:** Extremely positive, indicating high usability and integration, and showed no limiting barriers to effective system use. **Overall score: 100.**

The analysis of the SUS responses provided by the rescuers, none of whom had prior experience with AR, suggests that their lack of familiarity likely influenced their perceptions of the KHAIT system. Particularly, aspects of the system that were perceived as cumbersome could stem from the inherent learning curve associated with using AR technology for the first time. This novelty effect could have exaggerated feelings of the system being unintuitive or overly complex, as evidenced by varied responses regarding the need for technical support and the system's ease of use.

### 6.2.2 MRC Results.
The MRC questionnaire assesses users' apprehensions and concerns regarding the use of mixed reality systems, focusing on aspects such as security, privacy, social implications, and trust [30]. It is scored from 9 to 45, where higher scores indicate greater user apprehensions or concerns towards mixed reality systems. The scale interprets scores across a range: lower scores suggest minimal concerns, mid-range scores indicate moderate concerns, and higher scores reflect increased apprehensions about mixed reality technologies. Additionally, the last three questions of the questionnaire are reverse-scored (R), which means higher responses indicate lower concerns. Below is a summary of the responses and overall score by each participant in Figure 7:

All MRC questions used can be found in Appendix B. The average score among the rescuers was **21**, which indicated moderate concern for using mixed reality in the rescue trials. Rescuer 2 displayed the highest level of concern with a score of 24, indicating notable reservations about privacy, social implications, and trust aspects of KHAIT. In contrast, Rescuer 5 showed the least apprehension, with the lowest score of 16, suggesting a higher level of comfort and readiness to adopt KHAIT for SAR. Other rescuers' scores fell between these two extremes: Rescuer 1 and Rescuer 4 both scored 21, reflecting moderate concerns, while Rescuer 3 scored 23, slightly higher but still indicative of apprehensions.

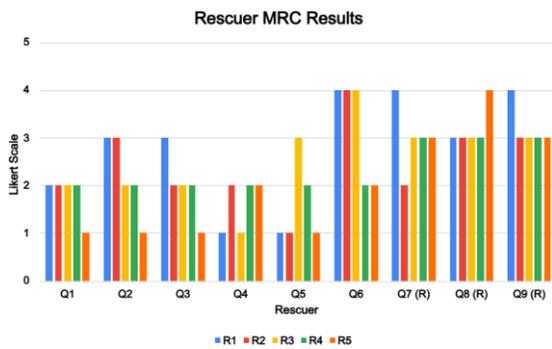

**Figure 7: Rescuer MRC Results**

The usability findings highlight the importance of a well-designed interface in high-stress environments, where ease of use and quick access to information can impact the efficiency and outcome of SAR operations. They also indicate the need to implement enhanced privacy protections and provide additional detailed training, which could mitigate apprehensions and improve overall user acceptance, allowing users to better appreciate mixed reality's benefits.

### 6.3 Canine Interaction with KHAIT

The rubble pile trials were designed to mimic USAR certification scenarios, where canines are required to navigate independently across rubble, often out of the handler's direct line of sight. This setup reflects the challenges faced in actual USAR environments. Throughout all the trials, we closely monitored the behavioral responses of the canines to the environments in both conditions. Observations were made by the handler and observers located on the pile and included the willingness to work independently (average approximate range from handler without requiring assistance), willingness to jump up (height and latency of jump) onto elevated surfaces, jump across voids (distance and latency), and latency of TFR. Only one dog appeared to habituate to the harness away from the pile but had a greater reduction in independence while working. In all other trials, the canines had no observable difference in any of the parameters between conditions.

An additional, unanticipated benefit of the KHAIT system was discovered during the rubble pile trials when a canine struggled to interpret the concealed subject's olfactory cues. When this occurred, the handlers could interpret that the canine was "working an odor of a subject" noting that the canines were either confused or struggling to pinpoint the odor location. This was observed via the HoloLens-2 when the canine would circle an area repeatedly. Without this information, the handler would be unaware that the dog could not pinpoint the area, delaying a response to the area and the canine could leave the subject without emitting its TFR, thus creating a false negative response. In these cases where no TFR was emitted, the handler was able to interpret the canine's behavior as focused interest on a particular area, in conjunction with the topography of the pile, to estimate where the subject might be. In a deployment scenario, these details would direct other resources, such as search cameras or acoustic monitoring devices, to the area.

## 7 Discussion

This section discusses aspects of the KHAIT system used in SAR operations. We revisit our research questions around (1) user experience in AR and (2) localization solutions for canine harnesses. We also discuss broader implications regarding the integration of HCI and ACI and the technological mediation of search dynamics. Finally, we discuss limitations and future work.

### 7.1 RQ1: User Experience in Augmented Reality

Addressing our first research question regarding the impact of AR on SAR operations, our study highlights how the integration of the KHAIT system, featuring a smart canine harness linked to a wearable SLAM-based AR headset for handlers, enhances operational effectiveness. KHAIT allows SAR teams to visualize real-time object detections and video feeds directly from a search dog's perspective. Such unique capabilities may assist in complex search environments where direct visual contact with the dog may not always be possible.

A unique aspect of KHAIT is its ability to project the harness's data, including the canine's location and environmental cues it encounters, directly into the handler's field of vision through the AR headset. The object detection model used, YOLOv8l, recognized details such as water bottles and other personal items that indicated where a person may have been. This direct data transmission aids handlers in making immediate, informed decisions by enhancing their situational awareness. As initially observed from our evaluation studies and observational feedback, key findings demonstrated optimized search strategies by rescuers and reduced time to locate survivors.

However, despite these benefits, our study also uncovered challenges related to the AR system's usability. Participants with limited prior exposure to AR technology experienced difficulties manipulating interface elements, such as adjusting video feeds, which introduced delays and increased cognitive load during operations. This challenge is often consistent in AR user studies with participants new to the technology and can be mitigated over time as the technology becomes more ubiquitous to society [45].

### 7.2 RQ2: Localization Solutions for Canine Harnesses

Concerning our second research question regarding the usability of wearable technology on SAR canines, our exploration into localization solutions revealed helpful ergonomic and behavioral considerations. Initially, we considered integrating SLAM systems into the canine harness for this purpose. However, our experiments and research have prompted several reflections on this ACI design. Firstly, SLAM systems require the fusion of multiple sensors and substantial computing power, as demonstrated by devices like the HoloLens-2, which integrates IMU sensors, stereo cameras, depth cameras, and powerful processing capabilities. While feasible for devices like the HoloLens-2, this approach would likely be impractical for animals due to the added weight and complexity [43]. Secondly, SLAM systems, primarily designed for robots, face challenges when adapting to the unique characteristics of animals, such as their high-speed movements and non-rigid body structures. This makes localization difficult in scenarios involving animals [13]. We explored more suitable localization solutions tailored to different environments as an alternative. For outdoor applications like wilderness SAR, GPS technology provides sufficiently accurate localization. In contrast, USAR operations can benefit from integrating lightweight edge devices with distributed systems like KHAIT. This system integration approach requires minimal equipment attached to animals, reducing their burden and aligning with HITL methods for effective and practical localization solutions in complex scenarios [13].

Another consideration in the design of KHAIT, which would similarly impact any SLAM system integration, concerns power management for computational requirements. The initial design of the KHAIT harness included a large mobile power bank to ensure sufficient power for prolonged operational use. This component was necessary to efficiently support the energy demands of the harness's computing and camera systems. Specifically, the NVIDIA Orin Nano, central to our system to run the computationally intensive YOLOv8l model, required up to 20V and a power consumption of 15W. If SLAM technologies were to be incorporated, this power source might prove inadequate due to the increased energy demand from additional sensors.

In resource-poor search environments, the operational duration of the KHAIT system could require backup power supplies to maintain functionality, particularly in extended missions. During our trials, which took place in high temperatures around 33˚C, we observed that excessive heat could cause AR headsets to overheat and shut down, necessitating the use of backup units to continue the evaluations without interruption.

Due to their complexity and weight, SLAM systems may offer robust localization capabilities but may not yet be suited for ACI applications. Exploring lightweight, alternative localization solutions such as Visual RGB and RGB-D sensory inputs tailored to the unique needs of animals presents a promising avenue for future research and development [64].

## 7.3 Broader Implications

*7.3.1 Bridging HCI and ACI.* Integrating AR and localization technologies into SAR operations involves integrating complex systems to accommodate animals' unique behavioral patterns and communication methods for enhancing collaboration between human rescuers and their canine counterparts. By leveraging wearable devices and distributed sensor networks, we enable real-time communication and reliable coordination from a HITL, boosting the effectiveness of rescue operations. These innovations allow the immediate translation of canine discoveries into actionable data for human rescuers, optimizing collective efforts and assuring coordinated emergency response.

A key aspect of our study was initially measuring these technologies' impact on operational efficiency. The 22% decrease in time needed to locate survivors with KHAIT is notable in the context of SAR operations. In disaster scenarios, where conditions can rapidly deteriorate, reducing response times by nearly a quarter can critically increase the chances of survival for victims [35, 44]. This time reduction not only speeds up the rescue process but also minimizes the exposure of rescuers and victims to potentially hazardous conditions, thus enhancing overall safety [8].

*7.3.2 Technological Mediation of Search Dynamics.* In USAR operations, where environments are often too unstable for rescuers to navigate freely, integrating the KHAIT system with SAR canines can enhance situational awareness. This technology enables handlers to gain real-time insight into the dynamic and challenging conditions within which their canines operate.

The technology mediated the canine-searcher interaction by replacing traditional follow-back methods with a visual path displayed directly on the AR interface. The path enabled the searcher to navigate quickly and accurately to the survivor's location, bypassing obstacles and potentially hazardous areas. This method expedited the search process, reducing traditional risks and inefficiencies associated with physical scouting. Upon reaching the source of the subject odor, the system can relay detailed visualizations of the debris, including its type, orientation, and stability, to other trained rescuers or display this information on external monitors for additional coordination. This could allow rescue teams to appropriately anticipate the resources needed to extricate the subject from their entombed location within the rubble.

Interestingly, the ability to observe, from the perspective of the USAR canine, the environment that they are engaging with gave the handler the ability to redirect or assist the canine before the canine becomes stressed or shut down. During the study, handlers sometimes gave assisting cues to help the canine traverse a particular area that was not obvious to the canine. For example, a canine might not recognize obstacles as being something they can climb or cross, but with the verbal cue from the handler, the canine was able to interact with that obstacle successfully.

## 7.4 Limitations and Future Work

One of the primary limitations of the current KHAIT system evaluation is the limited scope of the user and animal studies conducted. With nine participants, four of which acted as rescuers and two trained canines that could use the KHAIT harness during the trials, this small sample size means that the initial findings, while promising, primarily serve as preliminary evidence of success from our prototype testing. These results highlight the system's potential but cannot yet establish its efficacy across several SAR scenarios and conditions. Recognizing this, we have planned more rigorous IRB-approved and IACUC-approved studies involving a larger pool of SAR professionals and multiple canines at official USAR training facilities. As previously mentioned in Section 3.4, KHAIT integrates with Ajna's sensemaking capabilities, serving as a supplementary fail-safe that incorporates multiple HITL inputs to validate detections and foster swift consensus among a team of rescuers. The future studies will evaluate how both systems collaboratively enhance quick decision-making through consensus building, especially in rapid judgment scenarios. The studies will also include behavioral coding, allowing for direct analysis of the various components of a canine's search sequence.

The ergonomic design of the canine harness and the system's reliance on sophisticated technology present practical challenges. Currently, the harness is loose-fitting, and the weight is not evenly distributed, which, over longer trial periods, may hinder the canine's movement, increase fatigue, and inhibit full range of motion, reducing confidence in crossing unfamiliar rubble. To address these issues, we aim to continue developing prototype harnesses that securely house all components and evenly distribute weight above the forelimbs, preventing adverse pressure on the spine during movement. Items that may catch or become loose will be encased within the harness and isolated from environmental hazards.

Weight is a constant concern for USAR dogs, which are typically in the 15–35 kg range, and any harness should not exceed 10 percent of the dog's body weight to protect joints and the spine from injury. Future iterations of the KHAIT system will incorporate different-sized harnesses to accommodate the various sizes and body shapes of dogs while minimizing the bulk and overall coverage of the harness to reduce discomfort and maintain a stable platform for the system. Recognizing that a harness-worn system adds weight, we will minimize the weight of all components, ensure a snug fit, encase exothermic components in ventilated exterior pockets, and use heat-reflective material on the interior. This approach will also ensure that dogs can be systematically physically conditioned and acclimated to carry the extra weight without negative side effects.

## 8 Conclusion

Our findings highlight the value of integrating AR and advanced localization technologies in SAR operations, which aids human rescuers by enhancing their ability to locate and assist survivors. While the results indicated variable outcomes, on average, KHAIT demonstrated a reduced time for rescuers to reach survivors, suggesting its potential in improving operational efficiency in certain contexts. In practical terms, the enhanced situational awareness provided by AR and the precise tracking enabled by localization technologies may transform SAR efforts, making them faster, safer, and more coordinated. The 22% time reduction in survivor location we observed from initial USAR rescuers and canine trials indicate a promising path forward that could improve outcomes in real-world rescue scenarios, potentially saving lives.

## Acknowledgments

This research would not have been possible without funding assistance from the U.S. Army Combat Capabilities Development Command. We also recognize the willingness and generosity of Search and Rescue Assist Inc. (SARA), a Virginia non-profit organization that partnered with us and gave access to their official SAR training facilities. Additionally, we thank Yingjie Wang, Patrick Shih, the A3 Lab at Virginia Tech (https://ai.bse.vt.edu/), and the Crowd Intelligence Lab at Virginia Tech (https://crowd.cs.vt.edu/).

## A  System Usability Scale (SUS) Questions

(1) I think that I would like to use this system frequently.

(2) I found the system unnecessarily complex.

(3) I thought the system was easy to use.

(4) I think that I would need the support of a technical person to be able to use this system.

(5) I found the various functions in this system were well integrated.

(6) I thought there was too much inconsistency in this system.

(7) I would imagine that most people would learn to use this system very quickly.

(8) I found the system very cumbersome to use.

(9) I felt very confident using the system.

(10) I needed to learn a lot of things before I could get going with this system.

## B  Mixed Reality Concerns (MRC) Questions

(1) I am concerned about the possibility of non-authenticated individuals gaining access to this MR system.

(2) I am concerned about the potential exposure of sensitive data through this MR system to unauthorized parties.

(3) I worry that using this MR system might lead to my personal information being misused.

(4) I fear that with this MR system, it becomes increasingly hard to maintain a clear distinction between virtual behavior and real-life behavior.

(5) I am concerned about the potential of this MR system to influence my behaviors in ways that could be detrimental to my well-being.

(6) Using this MR system might make me appear disconnected from others in my physical environment.

(7) I believe that only legitimate individuals can access this MR system. **(R)**

(8) I am sure that this MR system is maintaining a secure environment. **(R)**

(9) I am confident that my anonymity is protected by this MR system. **(R)**

Items marked with an **(R)** are reverse-scored, meaning agreement indicates a positive perception of trust and security.